\RequirePackage{lineno}
\documentclass[aps,prl,showpacs,twocolumn,superscriptaddress,groupedaddress,floatfix,amsfonts]{revtex4}
\usepackage{graphicx}
\usepackage{dcolumn}
\usepackage{bm}
\usepackage{amssymb}
\usepackage{amsmath}
\usepackage{xspace}
\newcommand{\la}{\langle}
\newcommand{\ra}{\rangle}
\newcommand{\lt}{\!<\!}
\newcommand{\gt}{\!>\!}

\newcommand{\GeV}{\ensuremath{\text{GeV}}\xspace}
\newcommand{\TeV}{\ensuremath{\text{TeV}}\xspace}
\newcommand{\Ptg}{p_{T}^{\gamma}\xspace}
\newcommand{\ptg}{\ensuremath{p_T^{\gamma}}\xspace}
\newcommand{\gb}{\ensuremath{\gamma+{b}}\xspace}
\newcommand{\gc}{\ensuremath{\gamma+{c}}\xspace}

\usepackage{color}

\begin{document}

\hspace{5.2in} \mbox{FERMILAB-PUB-12-574-E}

\title{\boldmath Measurement of the differential $\gamma+c$-jet 
cross section and the ratio of differential $\gamma+c$ and $\gamma+b$ cross sections in $p\bar{p}$ collisions at $\sqrt{s}=1.96~\TeV$}
\affiliation{LAFEX, Centro Brasileiro de Pesquisas F\'{i}sicas, Rio de Janeiro, Brazil}
\affiliation{Universidade do Estado do Rio de Janeiro, Rio de Janeiro, Brazil}
\affiliation{Universidade Federal do ABC, Santo Andr\'e, Brazil}
\affiliation{University of Science and Technology of China, Hefei, People's Republic of China}
\affiliation{Universidad de los Andes, Bogot\'a, Colombia}
\affiliation{Charles University, Faculty of Mathematics and Physics, Center for Particle Physics, Prague, Czech Republic}
\affiliation{Czech Technical University in Prague, Prague, Czech Republic}
\affiliation{Center for Particle Physics, Institute of Physics, Academy of Sciences of the Czech Republic, Prague, Czech Republic}
\affiliation{Universidad San Francisco de Quito, Quito, Ecuador}
\affiliation{LPC, Universit\'e Blaise Pascal, CNRS/IN2P3, Clermont, France}
\affiliation{LPSC, Universit\'e Joseph Fourier Grenoble 1, CNRS/IN2P3, Institut National Polytechnique de Grenoble, Grenoble, France}
\affiliation{CPPM, Aix-Marseille Universit\'e, CNRS/IN2P3, Marseille, France}
\affiliation{LAL, Universit\'e Paris-Sud, CNRS/IN2P3, Orsay, France}
\affiliation{LPNHE, Universit\'es Paris VI and VII, CNRS/IN2P3, Paris, France}
\affiliation{CEA, Irfu, SPP, Saclay, France}
\affiliation{IPHC, Universit\'e de Strasbourg, CNRS/IN2P3, Strasbourg, France}
\affiliation{IPNL, Universit\'e Lyon 1, CNRS/IN2P3, Villeurbanne, France and Universit\'e de Lyon, Lyon, France}
\affiliation{III. Physikalisches Institut A, RWTH Aachen University, Aachen, Germany}
\affiliation{Physikalisches Institut, Universit\"at Freiburg, Freiburg, Germany}
\affiliation{II. Physikalisches Institut, Georg-August-Universit\"at G\"ottingen, G\"ottingen, Germany}
\affiliation{Institut f\"ur Physik, Universit\"at Mainz, Mainz, Germany}
\affiliation{Ludwig-Maximilians-Universit\"at M\"unchen, M\"unchen, Germany}
\affiliation{Fachbereich Physik, Bergische Universit\"at Wuppertal, Wuppertal, Germany}
\affiliation{Panjab University, Chandigarh, India}
\affiliation{Delhi University, Delhi, India}
\affiliation{Tata Institute of Fundamental Research, Mumbai, India}
\affiliation{University College Dublin, Dublin, Ireland}
\affiliation{Korea Detector Laboratory, Korea University, Seoul, Korea}
\affiliation{CINVESTAV, Mexico City, Mexico}
\affiliation{Nikhef, Science Park, Amsterdam, the Netherlands}
\affiliation{Radboud University Nijmegen, Nijmegen, the Netherlands}
\affiliation{Joint Institute for Nuclear Research, Dubna, Russia}
\affiliation{Institute for Theoretical and Experimental Physics, Moscow, Russia}
\affiliation{Moscow State University, Moscow, Russia}
\affiliation{Institute for High Energy Physics, Protvino, Russia}
\affiliation{Petersburg Nuclear Physics Institute, St. Petersburg, Russia}
\affiliation{Instituci\'{o} Catalana de Recerca i Estudis Avan\c{c}ats (ICREA) and Institut de F\'{i}sica d'Altes Energies (IFAE), Barcelona, Spain}
\affiliation{Uppsala University, Uppsala, Sweden}
\affiliation{Taras Shevchenko National University of Kyiv, Kiev, Ukraine}
\affiliation{Lancaster University, Lancaster LA1 4YB, United Kingdom}
\affiliation{Imperial College London, London SW7 2AZ, United Kingdom}
\affiliation{The University of Manchester, Manchester M13 9PL, United Kingdom}
\affiliation{University of Arizona, Tucson, Arizona 85721, USA}
\affiliation{University of California Riverside, Riverside, California 92521, USA}
\affiliation{Florida State University, Tallahassee, Florida 32306, USA}
\affiliation{Fermi National Accelerator Laboratory, Batavia, Illinois 60510, USA}
\affiliation{University of Illinois at Chicago, Chicago, Illinois 60607, USA}
\affiliation{Northern Illinois University, DeKalb, Illinois 60115, USA}
\affiliation{Northwestern University, Evanston, Illinois 60208, USA}
\affiliation{Indiana University, Bloomington, Indiana 47405, USA}
\affiliation{Purdue University Calumet, Hammond, Indiana 46323, USA}
\affiliation{University of Notre Dame, Notre Dame, Indiana 46556, USA}
\affiliation{Iowa State University, Ames, Iowa 50011, USA}
\affiliation{University of Kansas, Lawrence, Kansas 66045, USA}
\affiliation{Kansas State University, Manhattan, Kansas 66506, USA}
\affiliation{Louisiana Tech University, Ruston, Louisiana 71272, USA}
\affiliation{Northeastern University, Boston, Massachusetts 02115, USA}
\affiliation{University of Michigan, Ann Arbor, Michigan 48109, USA}
\affiliation{Michigan State University, East Lansing, Michigan 48824, USA}
\affiliation{University of Mississippi, University, Mississippi 38677, USA}
\affiliation{University of Nebraska, Lincoln, Nebraska 68588, USA}
\affiliation{Rutgers University, Piscataway, New Jersey 08855, USA}
\affiliation{Princeton University, Princeton, New Jersey 08544, USA}
\affiliation{State University of New York, Buffalo, New York 14260, USA}
\affiliation{University of Rochester, Rochester, New York 14627, USA}
\affiliation{State University of New York, Stony Brook, New York 11794, USA}
\affiliation{Brookhaven National Laboratory, Upton, New York 11973, USA}
\affiliation{Langston University, Langston, Oklahoma 73050, USA}
\affiliation{University of Oklahoma, Norman, Oklahoma 73019, USA}
\affiliation{Oklahoma State University, Stillwater, Oklahoma 74078, USA}
\affiliation{Brown University, Providence, Rhode Island 02912, USA}
\affiliation{University of Texas, Arlington, Texas 76019, USA}
\affiliation{Southern Methodist University, Dallas, Texas 75275, USA}
\affiliation{Rice University, Houston, Texas 77005, USA}
\affiliation{University of Virginia, Charlottesville, Virginia 22904, USA}
\affiliation{University of Washington, Seattle, Washington 98195, USA}
\author{V.M.~Abazov} \affiliation{Joint Institute for Nuclear Research, Dubna, Russia}
\author{B.~Abbott} \affiliation{University of Oklahoma, Norman, Oklahoma 73019, USA}
\author{B.S.~Acharya} \affiliation{Tata Institute of Fundamental Research, Mumbai, India}
\author{M.~Adams} \affiliation{University of Illinois at Chicago, Chicago, Illinois 60607, USA}
\author{T.~Adams} \affiliation{Florida State University, Tallahassee, Florida 32306, USA}
\author{G.D.~Alexeev} \affiliation{Joint Institute for Nuclear Research, Dubna, Russia}
\author{G.~Alkhazov} \affiliation{Petersburg Nuclear Physics Institute, St. Petersburg, Russia}
\author{A.~Alton$^{a}$} \affiliation{University of Michigan, Ann Arbor, Michigan 48109, USA}
\author{A.~Askew} \affiliation{Florida State University, Tallahassee, Florida 32306, USA}
\author{S.~Atkins} \affiliation{Louisiana Tech University, Ruston, Louisiana 71272, USA}
% ...
%
%
\collaboration{The D0 Collaboration\footnote{with visitors from
$^{a}$Augustana College, Sioux Falls, SD, USA,
$^{b}$The University of Liverpool, Liverpool, UK,
$^{c}$UPIITA-IPN, Mexico City, Mexico,
$^{d}$DESY, Hamburg, Germany,
$^{e}$SLAC, Menlo Park, CA, USA,
$^{f}$University College London, London, UK,
$^{g}$Centro de Investigacion en Computacion - IPN, Mexico City, Mexico,
$^{h}$ECFM, Universidad Autonoma de Sinaloa, Culiac\'an, Mexico
and
$^{i}$Universidade Estadual Paulista, S\~ao Paulo, Brazil.
}} \noaffiliation
\vskip 0.25cm

\date{\today}

\begin{abstract}
  We present measurements of the differential cross section ${\rm d}\sigma/{\rm d}\Ptg$
  for the associated production of a $c$-quark jet and an isolated photon
  with rapidity $|y^\gamma|\lt 1.0$ and  
  transverse momentum $30<\ptg <300$~\GeV. 
  The $c$-quark jets are required to have $|y^\text{jet}|\lt 1.5$ and $p_T^{\text {jet}}>15$ GeV.
  The ratio of differential cross sections for $\gamma+c$ to $\gamma+b$ production
  as a function  of $\Ptg$ is also presented.
  The results are based on data corresponding to an integrated luminosity of 8.7 fb$^{-1}$
  recorded with the D0 detector at the Fermilab Tevatron $p\bar{p}$ Collider at $\sqrt{s}=$1.96~\TeV.
  The obtained results are compared to next-to-leading order perturbative QCD calculations 
  using various parton distribution functions, to predictions
  based on the $k_{\rm T}$-factorization approach, and to predictions from the {\sc sherpa} and {\sc pythia} Monte Carlo event generators.
\end{abstract}

\maketitle

\linenumbers


In hadron-hadron collisions high-energy photons 
are mainly produced directly in a hard parton scattering process.
For this reason, and due to their pointlike electromagnetic coupling to the quarks,
they provide a clean probe of parton-level dynamics. 
Photons in association with a charm ($c$) quark are produced
primarily through the Compton-like scattering process
$gc\to \gamma c$, which dominates up to photon transverse momenta with respect to the
beam axis of $\Ptg \approx 70-80$~GeV, and through quark-antiquark annihilation, 
$q\bar{q}\to \gamma g \to \gamma c\bar{c}$, which dominates at higher $\Ptg$~\cite{Tzvet}.
Inclusive $\gc$ production may also 
originate from processes like $gg \to c\bar{c}$ or $cg \to cg$,
where the fragmentation of a final state $c$-quark or gluon produces a photon~\cite{Tzvet}.
Photon isolation requirements 
substantially reduce the contributions from these processes.
Measurements of the $\gc$-quark jet differential cross section as a function
of $\ptg$ 
improve our  understanding of the underlying production mechanism and  
provide useful input for the $c$-quark parton distribution functions (PDFs)  
of the colliding hadrons.

In this Letter, we present measurements of the inclusive $\gamma+c$-jet
production cross sections using data collected from June 2006 to September 2011 with the
D0 detector in $p\bar{p}$ collisions at $\sqrt{s}=1.96~\TeV$ which 
correspond to an integrated luminosity of $8.7 \pm 0.5$~fb$^{-1}$~\cite{d0lumi}. 
The cross section is measured differentially as a function of
$\Ptg$ for photons within rapidities $|y^\gamma|\lt 1.0$ and $30<\Ptg< 300$~GeV, while
the $c$-jet is required to have $|y^\text{jet}|\lt 1.5$ and $p_{T}^{\rm jet}>15$~GeV.
In comparison to our previous \gc-jet measurement~\cite{gamma_b_d0},
we now retain all events having at least one jet originating from a charm quark,
as opposed to considering only the events in which the charm jet candidate is the jet 
with highest $p_T$.
To increase the signal yield and study a trend 
in the data/theory ratio observed in Ref.~\cite{gamma_b_d0}, we have extended 
the rapidity~\cite{Rap} region from $| y^\text{jet}|\lt 0.8$ to $| y^\text{jet}|\lt 1.5$
and combine regions with positive and negative products of rapidities, $y^{\gamma}y^{\rm jet}$.
In addition, an increased integrated luminosity by about a factor of nine allows the
$\Ptg$ range to be extended to higher values.

The data set and event selections used in our measurement
are similar to those used in the recently published 
measurement of the $\gamma+b$-jet differential cross section \cite{gamma_b_d0_new}.
However, because of the difficulty in discriminating $c$ jets from light jets,
this measurement adopts a different strategy for the estimation of the $c$-jet fraction.
Here we apply a significantly more stringent requirement for selecting heavy flavor jets (originating from $c$ and $b$ quarks)
in order to suppress the rates of light jets (originating from light quarks or gluons) 
by an additional factor of $2.5-3$. This small residual contribution of light jets is then subtracted from the
selected data events prior to performing the fit 
with the discriminant templates of $b$-jets and $c$-jets to extract the $c$-jet fraction.
Using this event selection criteria, we reproduce the results
for the $\gamma+b$-jet cross section, measure the $\gamma+c$-jet  cross section
and calculate the ratio $\sigma(\gamma+c)/\sigma(\gamma+b)$  in bins of $\Ptg$.
Common experimental uncertainties and dependence 
on the higher-order corrections in theory are reduced in the ratio, allowing
a precise study of the 
relative $\sigma(\gamma+c)/\sigma(\gamma+b)$ rates.

The D0 detector is a general purpose detector 
described in detail elsewhere~\cite{d0det}.
The subdetectors most relevant to this analysis are the central tracking
system, composed of a silicon microstrip tracker (SMT) and a central fiber
tracker (CFT) embedded in a 1.9~T solenoidal magnetic field, the central
preshower detector (CPS), and the calorimeter.
The CPS is located immediately before the inner layer of the central calorimeter
and is formed of approximately one radiation length of lead absorber followed by three
layers of scintillating strips. The calorimeter consists of a central section (CC) with
coverage in pseudorapidity of $|\eta_{\rm det}|<1.1$~\cite{d0_coordinate}, 
and two end calorimeters (EC) extending coverage to $|\eta_{\rm det}| \approx 4.2$, all housed
in separate cryostats, with scintillators between the CC and EC cryostats providing sampling of
developing showers for $1.1 \lt |\eta_{\rm det}| \lt 1.4$.
The electromagnetic (EM) section of the
calorimeter is segmented longitudinally into four layers (EM$i$, $i=1-4$), 
with transverse segmentation into cells of size 
$\Delta\eta_{\rm det}\times\Delta\phi = 0.1\times 0.1$~\cite{d0_coordinate}, 
except EM3 (near the EM shower maximum), where it is $0.05\times 0.05$.
The calorimeter allows for a precise measurement of the energy
and direction of electrons and photons,
providing an energy resolution of approximately $4\%$~($3\%$) at an energy of $30~(100)$~GeV,
and an angular resolution of  about $0.01$ radians.
The energy response of the calorimeter to photons is calibrated using
electrons from $Z$ boson decays. Since electrons and photons interact
differently in the detector material before the calorimeter, additional energy corrections 
as a function of $\Ptg$ are derived using a detailed {\sc geant}-based~\cite{Geant} simulation 
of the D0 detector response. These corrections are largest, $\approx 2$\%,
at photon energies of about $30$ GeV.

The data used in this analysis  
are collected using a combination of triggers 
requiring a cluster of energy in the EM calorimeter with
loose shower shape requirements.
The trigger efficiency is $\approx\!96\%$ for photon candidates with 
$p_T^\gamma \approx\!\!30$~GeV and $\approx\!100\%$ for $p_T^\gamma>40$~GeV.

Offline event selection requires a reconstructed $p\bar{p}$ interaction vertex~\cite{pv} 
within 60~cm of the center of the detector along the beam axis. 
The missing transverse momentum in the event is required to be less than $0.7p_{T}^{\gamma}$ 
to suppress the background contribution from 
$W\to e\nu$ decays. These requirements are highly efficient ($\geq 98\%$) for signal events.

The photon selection criteria in the current measurement are identical to those used in Ref.~\cite{gamma_b_d0_new}.
The photon selection efficiency and acceptance
are calculated using samples of $\gamma+c$-jet events,
generated using the {\sc sherpa}~\cite{Sherpa} and {\sc pythia} \cite{PYT} event generators. 
The samples are processed through a {\sc geant}-based~\cite{Geant} 
simulation of the D0 detector response, followed by reconstruction using the same algorithms as applied to data.
As in Ref.~\cite{gamma_b_d0_new},
in the  efficiency and acceptance calculations the photon is required to be isolated at the particle level
by $E_T^{\rm iso} = E_T^{\rm tot}(0.4) - E_T^\gamma < 2.5$ GeV,
where $E_T^{\rm tot}(0.4)$ is the total transverse energy of particles within a cone of radius ${\cal R} = \sqrt{(\Delta\eta)^2+(\Delta\phi)^2} = 0.4$
centered on the photon and $E_T^\gamma$ is the photon transverse energy.
The particle level includes all stable particles as defined in Ref.~\cite{particle}.
The photon acceptance varies within ($82-90$)\% with a relative systematic uncertainty of ($2 - 5$)\%,
while the efficiency to pass photon identification criteria is ($68-85$)\% with 3\%
 systematic uncertainty.

At least one jet with $p_{T}^{\rm jet}>15$~GeV and $| y^\text{jet}|\lt 1.5$ 
must be reconstructed in each event. Jets are reconstructed
using the D0 Run~II algorithm~\cite{Run2Cone} with a cone radius of $\mathcal{R}=0.5$.  
The jet acceptance with respect to the $p_{T}^{\rm jet}$ and $|y^\text{jet}|$ 
varies between $91\%$ and $100\%$ in different $p_T^\gamma$ bins.
Uncertainties on the acceptance due to the jet energy scale, jet energy
resolution, and the difference  
between results obtained with {\sc sherpa} and {\sc pythia} are in the range of $(1-4)\%$. 
A set of criteria is imposed to have sufficient information to 
classify the jet as a heavy-flavor candidate:
the jet is required to have at least two associated tracks with $p_T>0.5$~\GeV
with at least one hit in the SMT, and at least one of these tracks must have $p_T>1.0$~\GeV.
These criteria have an efficiency of about 90\%.

To enrich the sample with heavy-flavor jets, a neural net based $b$-tagging algorithm ($b$-NN) 
is applied. It exploits the longer lifetimes of $b$-flavored hadrons in comparison to their lighter 
counterparts, after the rejection of long-lived $K^0_s$ and 
$\Lambda$ decays~\cite{b-NIM}. The inputs to the $b$-NN combine information from the impact parameter of 
displaced tracks and the topological properties of secondary vertices reconstructed in the jet to provide 
a continuous output value that tends towards one for $b$ jets and zero for light-quark jets. 
Events are required to contain at least one jet satisfying $b$-NN output $>0.7$.
This $b$-tagging selection suppresses light jets to less than $5\%$ of the heavy-flavor enhanced sample.
The efficiency for $b$ and $c$ jets to satisfy the $b$-tagging requirements in the simulation
is scaled by the data-to-Monte Carlo (MC) correction factors parametrized as a function of jet $p_T$ and $\eta$~\cite{b-NIM}.
Depending on $\ptg$, the selection efficiency for this requirement is ($8-10$)\% for $c$-jets
with relative systematic uncertainties of ($6-23$)\%, caused
by uncertainty on the data-to-MC correction factors.
The maximum difference between the efficiencies for $c$-jets
arising from the Compton-like and annihilation subprocesses is about 10\%.

The relative rate of remaining light jets (``light/all'') in the sample after the final selection is estimated using
{\sc sherpa} and {\sc pythia} $\gamma+$jet events, 
taking into account the data-to-MC correction factors as described in Ref.~\cite{b-NIM}.
The light jet rates predicted by {\sc pythia} and {\sc sherpa} agree within 5\%.
The central predictions are taken from {\sc sherpa}, which agrees with measured 
$\gamma+$jet \cite{SHERPA_gam,D0_gj} and $\gamma+b$-jet \cite{gamma_b_d0_new} cross sections within $(10-25)\%$.

After application of all selection requirements, 130,875 events remain.  
We estimate the photon purity using an artificial neural network discriminant~\cite{gamma_b_d0_new}.
The distribution of the output of this discriminant ($O_{\rm NN}$) is fitted to a
linear combination of templates for photons and jets obtained from
simulated $\gamma~+$ jet and dijet samples, respectively.  An
independent fit is performed in each $\Ptg$ bin. It yields photon
purities between 62\% and 99\%,
which are close to those obtained in Ref.~\cite{gamma_b_d0_new}.
Their systematic uncertainties are of a comparable magnitude, (5--9)\%. 

The invariant mass of all charged particles associated with a displaced secondary vertex in a jet, $M_{\rm SV}$,
is a powerful variable to discriminate $c$ from $b$ jets.
Since the $M_{\rm SV}$ templates for light and $c$-jets after
application of tight b-tagging requirements are quite close to each other,
we first subtract the remaining small fraction (1 -- 5\%) of light jets from the data.
Then the $c$-jet fraction is determined by fitting 
$M_{\rm SV}$ templates for $c$ and $b$ jets to the ($\gamma$+heavy flavor jet) data.
Jets from $b$ quarks contain secondary vertices that have in general larger values of $M_{\rm SV}$ as compared
to $c$ jets and the region beyond $M_{\rm SV}>2.0$ GeV is strongly dominated by $b$ jets.
The templates for $b$ and $c$ jets are obtained from {\sc pythia} samples of 
$\gamma+b$-jet and $\gamma+c$-jet events, respectively, and are
consistent with the templates generated using {\sc sherpa}.
The templates for jets arising from the Compton-like and annihilation 
subprocesses are also similar to each other.

\begin{figure}
\includegraphics[width=0.85\linewidth]{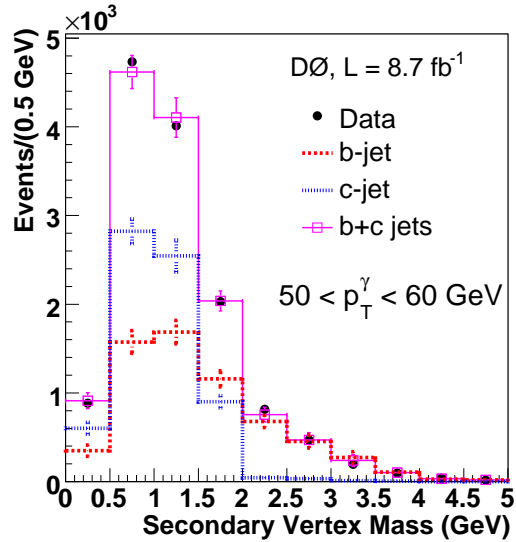} 
\caption{(Color online) Distribution of secondary vertex mass
  after all selection criteria for a representative bin of $50<\Ptg< 60$~GeV.  
  The expected contribution from the light-jet component has been subtracted from the data.
  The distributions for the $b$-jet and $c$-jet templates (with statistical uncertainties) 
  are shown normalized to their respective fitted fractions.}
\label{fig:cbjet_test}
\end{figure}

\begin{figure}
\includegraphics[width=0.9\linewidth]{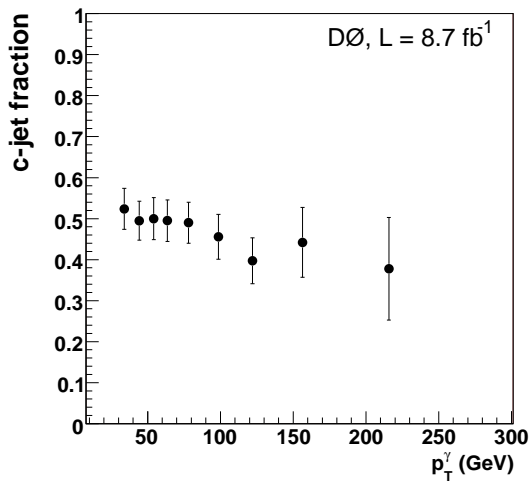} 
~\\[-4mm]
\caption{The $c$-jet fraction in data after subtraction of light-jet background  
as a function of $\Ptg$ derived from the template fit to the heavy quark jet data sample
after applying all selections. The error bars include statistical and systematical uncertainties.
Binning is the same as given in Table \ref{tab:xsect_cc}.}
\label{fig:b_pur_cc}
\end{figure}

The result of a maximum likelihood fit to the $M_{\rm SV}$ templates, normalized to the number of
events in data, is shown in Fig.~\ref{fig:cbjet_test} for the $50<\Ptg<60~\GeV$  bin as an example.  
Fits in the other \ptg bins are of similar quality.
As shown in Fig.~\ref{fig:b_pur_cc},
the estimated $c$-jet fraction obtained from the fits in the final selected heavy-flavor sample 
after subtraction of the light-jet component drops with increasing $\Ptg$, 
on average,  from about 52\% to about 40\%. 
The corresponding fit uncertainties range between
(4--32)\%, increasing towards higher $\Ptg$, and are dominated by the limited data statistics.
Since the fits are performed independently in each $\Ptg$ bin, these uncertainties are uncorrelated from bin to bin.
Additional systematic uncertainties are estimated by varying the
relative rate of light jets
by  $\pm50\%$  and by considering the differences in the light jet predictions from
{\sc sherpa} and {\sc pythia} event generators.
These two sources lead to uncertainties on the $c$-jet fraction of about (5--9)\% and 6\%, respectively.

Systematic uncertainty on the measured cross sections due to the $b$-NN selection is estimated by 
performing the measurement with looser $b$-NN selections: requiring $b$-NN output $>0.3$ or $>0.5$ instead of 0.7.
In both cases, this significantly increases the light-jet rate and also changes the $c$- and $b$-jet fractions,
resulting in a variation of the $\gamma+c$-jet cross section of $\leq7\%$.
This variation is taken as a systematic uncertainty on the cross section.

The data, corrected for photon and jet acceptance, reconstruction efficiencies
and the admixture of background events, are presented at the particle level~\cite{particle}
for comparison with predictions by unfolding the data for effects of detector resolution.

\begin{figure}
\includegraphics[width=1.03\linewidth]{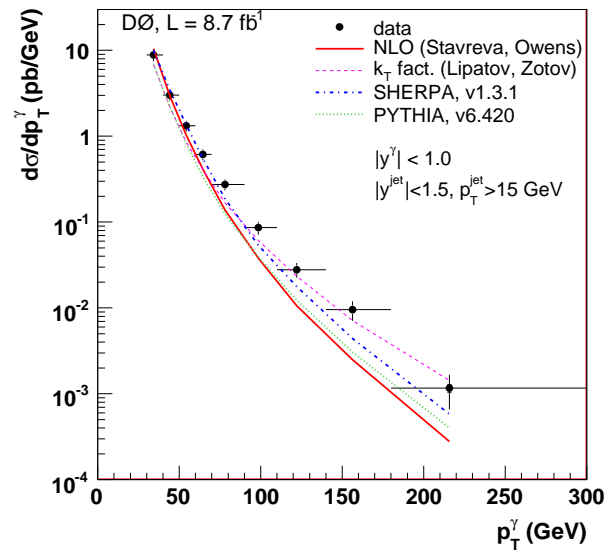}
~\\[-4mm]
\caption{(Color online) The \gc-jet differential production cross sections
  as a function of $\Ptg$.
 The uncertainties on the data points include statistical and systematic contributions added in quadrature. 
 The horizontal error bars show the $\Ptg$ bins.
 The measurements are compared to the NLO QCD calculations~\cite{Tzvet,Harris} using {\sc cteq}6.6M
  PDFs~\cite{CTEQ} (solid line). The predictions from {\sc sherpa}~\cite{Sherpa}, {\sc pythia}~\cite{PYT} and $k_{\rm T}$ factorization approach~\cite{Zotov,Zotov2} 
  are shown by the dash-dotted, dotted and dashed lines, respectively.}
\label{fig:xsectLOGplot}
\end{figure}

The differential cross sections of $\gamma+c$-jet production are extracted in nine bins of \ptg{}. 
They are listed in Table~\ref{tab:xsect_cc} and are shown in Fig.~\ref{fig:xsectLOGplot}. 
The data points are plotted at the values of $\Ptg$ for which
the value of a smooth function describing the dependence of the cross section on \ptg
equals the averaged cross section in the bin~\cite{TW}.

The statistical uncertainty of the results ranges from 2\% in the
first $\Ptg$ bin to $11\%$ in the last $\Ptg$ bin. The total
systematic uncertainty varies between 14\% and 42\% across these bins.  The main sources of 
uncertainty at low \ptg{} are due to the photon purity (up to $8\%$), the
$c$-jet fraction ($10-33\%$), and the luminosity (6\%)~\cite{d0lumi}.  
The total systematic uncertainties ($\delta_{\rm syst}$) and 
the bin-to-bin uncorrelated components ($\delta_{\rm syst}^{\rm unc}$) are shown
in Table~\ref{tab:xsect_cc}.

\begin{table*}
\centering
\caption{The \gc-jet production cross sections ${\rm d}\sigma/{\rm d}\Ptg$ in bins of $\Ptg$ for $|y^\gamma|<1.0$, $p_T^{\text {jet}}>15$ GeV and $|\eta^{\rm jet}|<1.5$
together with statistical uncertainties ($\delta_{\text{stat}}$),  total systematic uncertainties ($\delta_{\text{syst}}$), and the uncorrelated component of $\delta_{\text{syst}}$ 
($\delta_{\rm syst}^{\rm unc}$).
The column $\delta_{\rm tot}$ shows total experimental uncertainty obtained by adding $\delta_{\text{stat}}$ 
and $\delta_{\text{syst}}$ in quadrature.
The last four columns show theoretical predictions obtained within NLO QCD, $k_{\rm T}$-factorization, 
and by the {\sc pythia} and {\sc sherpa} event generators.}
\label{tab:xsect_cc}
\begin{tabular}{cccccccccc} \hline \hline
 ~$\Ptg$ bin~ & ~$\la\Ptg\ra$~ & \multicolumn{8}{c}{${\rm d}\sigma/{\rm d}\Ptg$ (pb/GeV) } \\\cline{3-10}
 (GeV) & (GeV) & Data & $\delta_{\rm stat}$($\%$) & $\delta_{\rm syst}(\delta_{\rm syst}^{\rm unc})$($\%$) & $\delta_{\rm tot}$($\%$) & ~~~NLO QCD~~~ & ~~~$k_{\rm T}$ fact.~~~ & ~~~{\sc pythia}~~~ & ~~~{\sc sherpa}~~~\\\hline  
   30 --  40 &  34.2 &  8.83 &    2 &   15~(3) &   15 &   10.5 &  6.88 &  6.55 &  10.0 \\\hline
   40 --  50 &  44.3 &  3.02 &    3 &   14~(3) &   15 &   2.96 &  2.19 &  2.21 &  3.47 \\\hline
   50 --  60 &  54.3 &  1.33 &    3 &   14~(4) &   14 &   1.03 &  8.59$\times 10^{-1}$ &  8.10$\times 10^{-1}$ &  1.36 \\\hline
   60 --  70 &  64.5 &  6.15$\times 10^{-1}$ &    3 &   14~(5) &   14 &   4.15$\times 10^{-1}$ &  4.12$\times 10^{-1}$ &  3.39$\times 10^{-1}$ &  5.52$\times 10^{-1}$ \\\hline
   70 --  90 &  78.1 &  2.73$\times 10^{-1}$ &    3 &   14~(5) &   14 &   1.39$\times 10^{-1}$ &  1.68$\times 10^{-1}$ &  1.24$\times 10^{-1}$ &  1.87$\times 10^{-1}$ \\\hline
   90 -- 110 &  98.6 &  8.61$\times 10^{-2}$ &    4 &   16~(8) &   17 &   3.80$\times 10^{-2}$ &  6.09$\times 10^{-2}$ &  3.90$\times 10^{-2}$ &  5.36$\times 10^{-2}$ \\\hline
  110 -- 140 & 122 &  2.79$\times 10^{-2}$ &    5 &   19~(11) &   19 &   1.06$\times 10^{-2}$ &  2.34$\times 10^{-2}$ &  1.23$\times 10^{-2}$ &  1.77$\times 10^{-2}$ \\\hline
  140 -- 180 & 156 &  9.54$\times 10^{-3}$ &    7 &   24~(17) &   26 &   2.49$\times 10^{-3}$ &  7.11$\times 10^{-3}$ &  3.07$\times 10^{-3}$ &  4.39$\times 10^{-3}$ \\\hline
  180 -- 300 & 216 &  1.16$\times 10^{-3}$ &   11 &   42~(32) &   43 &   2.79$\times 10^{-4}$ &  1.44$\times 10^{-3}$ &  4.01$\times 10^{-4}$ &  5.83$\times 10^{-4}$ \\\hline\hline
\end{tabular}
\end{table*}

Next-to-leading order (NLO) perturbative QCD 
predictions of order ${\cal O}({\alpha_{\rm s}^2})$~\cite{Tzvet,Harris}, 
with the renormalization scale $\mu_{R}$, factorization scale $\mu_{F}$,
and fragmentation scale $\mu_f$ all set to $\Ptg$, are given in Table \ref{tab:xsect_cc}.
The uncertainty from the scale choice is estimated 
through a simultaneous variation of all three scales by a factor of two, i.e.,
for $\mu_{R,F,f}=0.5 p_T^\gamma$ and $2 p_T^\gamma$,
and is found to be similar to those for $\gamma+b$-jet predictions ($5-30$)\%, being larger at higher $\Ptg$ \cite{gamma_b_d0_new}.
The NLO predictions utilize {\sc cteq}6.6M PDFs~\cite{CTEQ} and are corrected for non-perturbative effects
of parton-to-hadron fragmentation and multiple parton interactions. 
The latter are evaluated using {\sc sherpa} and {\sc pythia} MC samples generated using their default
settings~\cite {Sherpa,PYT}.
The overall corrections vary within $0.90-0.95$ 
with an uncertainty of $\lesssim 2\%$ assigned to account for the
difference between the two MC generators.

\begin{figure}
\includegraphics[width=1.03\linewidth]{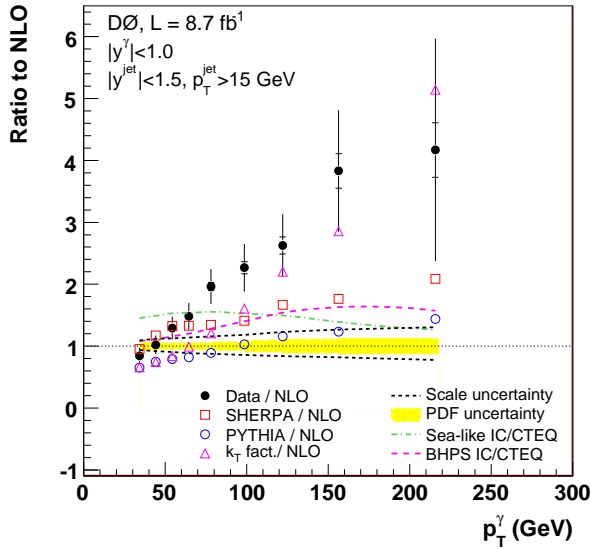}
~\\[-4mm]
\caption{(Color online) 
The ratio of \gc-jet production cross sections to NLO predictions for data and theoretical predictions.
The uncertainties on the data include both statistical (inner error bar) and total uncertainties (full error bar).  
Also shown are the uncertainties on the theoretical QCD scales and the {\sc cteq}6.6M PDFs. 
The ratio for intrinsic charm models \cite{CTEQc} are presented.
as well as the predictions given by $k_{\rm T}$-factorization \cite{Zotov,Zotov2}, {\sc sherpa}~\cite{Sherpa} and {\sc pythia}~\cite{PYT}.
}
\label{fig:xsectratio}
\end{figure}

The predictions based on the $k_{\rm T}$-factorization approach~\cite{Zotov,Zotov2} 
and unintegrated parton distributions~\cite{UPD}
are also given in Table \ref{tab:xsect_cc}.
The resummation of gluon diagrams with gluon transverse momentum ($k_{\rm T}$)
above a scale $\mu$ of order $1$~GeV,
leads to a broadening of the photon transverse momentum distribution 
in this approach~\cite{Zotov}.
The scale uncertainties on these predictions 
vary from about $-28\%/+31\%$ at $30<\Ptg<40~\GeV$ to about $+14\%/+5\%$ in the last $\Ptg$ bin.

Table \ref{tab:xsect_cc} also contains predictions from the {\sc pythia} \cite{PYT} event generator with the {\sc cteq}6.1L PDF set.
It includes only $2\to 2$ matrix elements (ME) with $gc\to \gamma c$ and $q\bar{q}\to \gamma g$ scatterings 
(defined at LO) followed by $g\to c\bar{c}$ splitting in the parton shower (PS).
We also provide predictions by the
{\sc sherpa} MC event generator \cite{Sherpa} with the {\sc cteq}6.6M  PDF set~\cite{CTEQ}.
Matching between the ME partons and the PS jets follows the
prescription given in Ref.~\cite{SHERPA_gam}, with the matching scale taken to be 15~GeV.
Systematic uncertainties are estimated by varying the ME-PS matching scale by $\pm 5$ GeV
around the chosen central value \cite{Sherpa_scale},
resulting in a $\pm7\%$ cross section variation.

All theoretical predictions are obtained using the photon isolation requirement 
of $E_T^{\rm iso}<2.5$ GeV.
The predictions are compared to data in Fig.~\ref{fig:xsectLOGplot} as a function of $\Ptg$.
The ratios of data over the NLO QCD calculations and of the various theoretical 
predictions to the NLO QCD calculations are presented in Fig.~\ref{fig:xsectratio}.
The NLO predictions with {\sc cteq}6.6M agree with 
{\sc mstw2008}~\cite{mstw} and {\sc abkm09nlo}~\cite{abkm} within 10\%.
Parameterizations for models containing intrinsic charm (IC) have been included 
in {\sc cteq6.6}c \cite{CTEQc}. 
Here we consider the BHPS IC model \cite{BHPS,BHPS2}, based on the Fock space picture of the nucleon structure \cite{Brod},
in which intrinsic charm appears mainly at large momentum fractions $x$,
and the sea-like model in which the charm PDF is sea-like, similar to that of the light-flavor sea quarks.
The NLO QCD predictions based on these intrinsic charm models are normalized to the standard {\sc cteq} 
predictions and are also shown in Fig.~\ref{fig:xsectratio}. 
Both non-perturbative intrinsic charm models predict a higher \gc-jet cross section. 
In the case of the BHPS model, the ratio grows with $\Ptg$, 
while an opposite trend is exhibited by the sea-like model.

The measured cross sections are in agreement with the NLO QCD predictions 
within theoretical and experimental uncertainties 
in the region of $30< \Ptg \lesssim 70$ GeV, but show systematic disagreement
for larger $\Ptg$. The cross section slope in data 
differs significantly from the NLO QCD prediction.
The results suggest a need for higher-order perturbative QCD corrections
in the large $\Ptg$ region, which is dominated by the annihilation process 
$q\bar{q}\to \gamma g$ (with $g\to c\bar{c}$),
and resummation of diagrams with additional gluon radiation. 
In addition, the underestimation of the rates for diagrams with $g\to c\bar{c}$  
splittings may result in lower theoretical predictions of cross sections
as suggested by LEP~\cite{LEP}, LHCb \cite{LHCb} and ATLAS \cite{Atlas} results.
The prediction from the $k_{\rm T}$-factorization approach is in better agreement with data
at $\Ptg\gt120$ GeV. However, it underestimates the cross section in the low and intermediate $\Ptg$ region.
The \gc-jet cross section as predicted by {\sc sherpa}  
becomes higher than the NLO QCD prediction at large $\Ptg$, but is still 
lower than the measured values.
It has been suggested that combining NLO parton-level calculations for the ME with PS predictions~\cite{NLOME}
will improve the description of the data~\cite{MStalk}.
\begin{table*}
\centering
\caption{The $\sigma(\gamma+c)/\sigma(\gamma+b)$ cross section ratio in bins of $\Ptg$ for $|y^\gamma|<1.0$, $p_T^{\text {jet}}>15$ GeV and $|\eta^{\rm jet}|<1.5$
together with statistical uncertainties ($\delta_{\text{stat}}$),  total systematic uncertainties ($\delta_{\text{syst}}$), and the uncorrelated component of $\delta_{\text{syst}}$ 
($\delta_{\rm syst}^{\rm unc}$).
The column $\delta_{\rm tot}$ shows total experimental uncertainty obtained by adding $\delta_{\text{stat}}$ 
and $\delta_{\text{syst}}$ in quadrature.
The last four columns show theoretical predictions obtained using NLO QCD, $k_{\rm T}$-factorization,
{\sc pythia} and {\sc sherpa} event generators.}
\label{tab:xsect_cb}
\begin{tabular}{cccccccccc} \hline\hline
 ~$\Ptg$ bin~ & ~$\la\Ptg\ra$~ & \multicolumn{8}{c}{$\sigma(\gamma+c)/\sigma(\gamma+b)$ } \\\cline{3-10}
  (GeV) & (GeV) & Data & $\delta_{\rm stat}$($\%$) & $\delta_{\rm syst}(\delta_{\rm syst}^{\rm unc})$($\%$) & $\delta_{\rm tot}$($\%$) & ~~~NLO QCD~~~ & ~~~$k_{\rm T}$ fact.~~~ & ~~~{\sc pythia}~~~ 
& ~~~{\sc sherpa}~~~\\\hline
   30 --  40 &  34.2 &   5.83 &    1 &    6~(3) &    6 &   5.81 &   4.30 &   5.10 &   6.17 \\\hline
   40 --  50 &  44.3 &   5.03 &    1 &    6~(3) &    6 &   5.28 &   4.01 &   4.97 &   5.28 \\\hline
   50 --  60 &  54.3 &   4.90 &    1 &    7~(3) &    7 &   4.79 &   3.83 &   4.66 &   4.79 \\\hline
   60 --  70 &  64.5 &   4.55 &    1 &    8~(4) &    8 &   4.37 &   3.91 &   4.34 &   4.21 \\\hline
   70 --  90 &  78.1 &   4.97 &    1 &    8~(4) &    8 &   3.83 &   3.88 &   3.99 &   3.54 \\\hline
   90 -- 110 &  98.6 &   4.22 &    2 &    9~(6) &    9 &   3.19 &   3.83 &   3.59 &   2.95 \\\hline
  110 -- 140 & 122 &   3.73 &    3 &   10~(6) &   11 &   2.60 &   3.86 &   3.00 &   2.50 \\\hline
  140 -- 180 & 156 &   4.34 &    5 &   13~(10) &   14 &   2.12 &   3.53 &   2.44 &   2.19 \\\hline
  180 -- 300 & 216 &   3.38 &    8 &   26~(22) &   27 &   1.73 &   4.04 &   1.98 &   1.93 \\\hline\hline
\end{tabular}
\end{table*}
\begin{figure}
\includegraphics[width=1.05\linewidth]{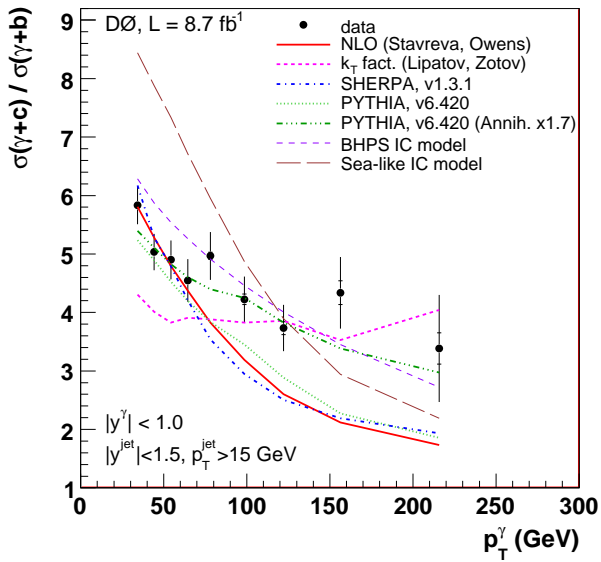}
~\\[-4mm]
\caption{(Color online)
The ratio of \gc-jet and \gb-jet production cross sections for data together with theoretical predictions as a function of $\Ptg$.
The uncertainties on the data include both statistical (inner error bar) and total uncertainties (full error bar).  
Predictions given by $k_{\rm T}$-factorization \cite{Zotov,Zotov2}, {\sc sherpa} 
\cite{Sherpa} and {\sc pythia} \cite{PYT} are also shown. The {\sc pythia} predictions with a contribution from the annihilation
process increased by a factor of 1.7 are shown as well.
The predictions for intrinsic charm models \cite{CTEQc} are also presented.
}
\label{fig:xsectratio2}
\end{figure}

In addition to measuring the \gc-jet cross-section,
we also obtain results for the $\gamma+b$-jet cross section using the new tight $b$-NN selection.
The values of the obtained $\gamma+b$-jet cross section
agree within 10\% (i.e. within uncertainties) with the published results [5]
obtained with a looser $b$-NN selection. 
We use them to calculate the ratio $\sigma(\gamma+c)/\sigma(\gamma+b)$ in bins of $\Ptg$.
In this ratio, many experimental systematic uncertainties cancel. 
Also, theory predictions of the ratio are less sensitive to the scale uncertainties,
and effects from missing higher-order terms that impact the normalizations of the cross sections.
The remaining uncertainties are caused by largely ($65-67\%$) correlated uncertainties 
coming from the fitting of $c$-jet and $b$-jet $M_{\rm SV}$ templates to data,
and by other uncertainties on the $c$-jet fractions discussed above.
The systematic uncertainties on the ratio vary within ($6-26$)\%, being largest at high $\Ptg$.
Theoretical scale uncertainties, estimated by
varying scales by a factor of two (to $\mu_{R,F,f}=0.5 p_T^\gamma$ and $2 p_T^\gamma$)
in the same way for $\sigma(\gamma+c)$ and $\sigma(\gamma+b)$  predictions,
are also significantly reduced.
Specifically, residual scale uncertainties are typically $\lesssim 10\%$ 
for the $k_{\rm T}$-factorization approach 
and $\lesssim 4\%$ for NLO QCD,
which indicates a much smaller dependence of the ratio on the higher-order corrections.
Experimental results as well as theoretical predictions for the ratios are presented in Table \ref{tab:xsect_cb}.

Figure \ref{fig:xsectratio2} shows the measured ratio $\sigma(\gamma+c)/\sigma(\gamma+b)$ 
as a function of $\Ptg$ and a comparison with various predictions.
There is good agreement with NLO QCD, {\sc sherpa} and {\sc pythia} predictions
in the region $30< \Ptg \lesssim 70$ GeV, 
while $k_{\rm T}$-factorization predicts smaller ratios than observed in data.
At higher $\Ptg$, data show systematically higher ratios than NLO QCD, 
{\sc sherpa} and {\sc pythia} predictions,
while $k_{\rm T}$-factorization starts agreeing with data within uncertainties.
We also show NLO predictions with the BHPS~\cite{BHPS,BHPS2} and sea-like IC models \cite{CTEQc} used
to predict \gc-jet cross section, while standard {\sc cteq6.6M} is used
to predict the \gb-jet cross section.
The BHPS model agrees with data at $\Ptg>80$ GeV, while the sea-like model
is significantly beyond the range of data points.
BHPS model would better describe the ratio to data with a small shift in normalization.
As with the \gc-jet measurement, the $\sigma(\gamma+c)/\sigma(\gamma+b)$ ratio can also be better described
by larger $g\rightarrow c\bar{c}$ rates than those used in the current 
NLO QCD, {\sc sherpa} and {\sc pythia} predictions.
To test this, we have increased the rate of the annihilation process  
(where $c$ jet is always produced due to $g \rightarrow c\bar c$ splitting) in the {\sc pythia} predictions. 
The best description of data is achieved  by increasing the rates by a factor of $1.7$
with $\chi^2/\rm {ndf} \simeq 0.7$ 
(compared to $\chi^2/\rm {ndf} = 4.1$ if such a factor is unity).
However, according to our estimates using the signal events simulated with {\sc sherpa}, 
there are also about (10--35)\% (higher for larger $p_T^\gamma$) events with two $c$-jets.
Assuming that one jet is coming from gluon initial state radiation followed by
$g\to c \bar c$ splitting, the required overall correction factor would be 
smaller by about (8--24)\%.

In conclusion, we have measured the differential cross section 
of $\gamma+c$-jet production as a function of $\Ptg$ at the Fermilab Tevatron $p\bar{p}$ collider. 
Our results cover the kinematic range $30<\ptg<300~\GeV$, $p_{T}^{\rm jet}>15$ GeV,
$|y^\gamma|<1.0$, and $|y^{\rm jet}|<1.5$. 
In the same kinematic region, and in the same $\Ptg$ bins, we have measured the $\sigma(\gamma+c)/\sigma(\gamma+b)$ 
cross section ratio.
None of the theoretical predictions considered give good description
of the data in all $\Ptg$ bins. Such a description 
might be achieved by including higher-order corrections
into the QCD predictions, while at $\Ptg\gtrsim 80$ GeV
the observed difference from data may also be
caused by an underestimated contribution from 
 gluon splitting $g\to c\bar{c}$ \cite{LEP,LHCb,Atlas} 
in the annihilation process or by contribution from intrinsic charm.
The presented results can be used for further development of theoretical models
to understand production of high energy photons in association with heavy flavor jets.

We are grateful to the authors of the theoretical calculations,
T.~Stavreva, J.~Owens, N.~Zotov, and F.~Siegert 
for providing dedicated predictions and for many useful discussions.

%
We thank the staffs at Fermilab and collaborating institutions,
and acknowledge support from the
DOE and NSF (USA);
CEA and CNRS/IN2P3 (France);
MON, NRC KI and RFBR (Russia);
CNPq, FAPERJ, FAPESP and FUNDUNESP (Brazil);
DAE and DST (India);
Colciencias (Colombia);
CONACyT (Mexico);
NRF (Korea);
FOM (The Netherlands);
STFC and the Royal Society (United Kingdom);
MSMT and GACR (Czech Republic);
BMBF and DFG (Germany);
SFI (Ireland);
The Swedish Research Council (Sweden);
and
CAS and CNSF (China).

\bibliography{paper_gc}

\begin{thebibliography}{99}
\bibitem{Rap}
The rapidity $y$ is related to the polar scattering 
angle $\theta$ with respect to the proton beam axis by
$y = 0.5\ln[(1 + \beta \cos\theta)/(1 - \beta\cos\theta)]$, 
where $\beta$ is defined as the ratio between momentum 
and energy $\beta = |\vec{p}|/E$.
\bibitem{d0_coordinate}
The polar angle $\theta$ and the azimuthal angle $\phi$ are defined
with respect to the positive $z$ axis, which is along the proton beam direction.
Pseudorapidity is defined as $\eta=-\ln[\tan(\theta/2)]$.
Also, $\eta_{\text {det}}$ is the pseudorapidity 
measured with respect to the center of the detector.
\bibitem{pv}
The primary $p\bar{p}$ interaction vertex is the most likely hard collision point,
among possibly several collisions within a specific beam crossing. The algorithm for defining primary vertex can be found in \cite{b-NIM}.
\bibitem{Sherpa_scale}
We  choose the following ME-PS matching parameters: the energy scale 
$Q_0 = 15$ GeV and parameter $D = 0.4$, where $D$ is taken to
be of the size of the photon isolation cone.
\end{thebibliography}
\bibliographystyle{apsrev}

\end{document}